\documentclass{article}
\usepackage{graphicx} 
\usepackage{natbib}
\usepackage{xcolor}

\usepackage[a4paper, total={6in, 8in}]{geometry}
\title{How different are deterministic physics suites when coupled to fixed model dynamics and why?}
\usepackage{xcolor}

\author{Edward Groot, Hannah Christensen, Xia Sun, Kathryn Newman, Wahiba Lfarh,\\ Romain Roehrig, Lisa Bengtsson, Julia Simonson
\\ 
Contact: large.edward.simulations@gmail.com \\edward.groot@physics.ox.ac.uk }
\date{January 2026}

\begin{document}
\maketitle
\begin{abstract}
   It is often difficult to attribute uncertainty and errors in atmospheric models to designated model components. This is because sub-grid parameterised processes 
   interact strongly with the large-scale transport represented by the explicit model dynamics. We carry out experiments with prescribed large-scale dynamics and different sub-grid physics suites. This dataset has been constructed for the Model Uncertainty Model Intercomparison Project (MUMIP), in which each suite forecasts sub-grid tendencies at a 22km grid. The common dynamics is derived from a convection-permitting benchmark: an ICON DYAMOND experiment (2.5km grid). 
   \\  
   We compare four different physics suites for atmospheric models in an Indian Ocean experiment. We analyse their joint PDFs of precipitation and associated physics tendencies for a full month. Precipitation is selected because it is a dominant uncertainty in the models that redistributes large amounts of heat. 
We find that all physics suites produce very similar precipitation amounts, with very high correlations between models, which exceed 0.95 at the 
native grid. 
However, the convection-permitting benchmark is more dissimilar from each of the physics suites, with correlations of $\approx$0.80. 
Similarly, we show that the vertically averaged physics tendencies in the free-troposphere are highly similar between the four physics suites, yet different if reconstructed for the benchmark. The water vapour sink is very closely linked with precipitation in the four physics suites. This suggests that the coarse-grid models are overconfident.\\ 
We hypothese is that variation in unresolved convective structures can lead to variation in the dynamics, following a given amount of latent heating at fine grids, but not in our physics suites. The difference is caused by the explicit interactions between gravity waves, which occurs at fine grids only. \\
We assess whether a non-linear feedback from convective precipitation systems explains our joint PDFs of precipitation. A slightly exponential highlights the 
significance of gravity-wave interactions. 
\\
These findings are further evidence for a non-linear feedback between convective organisation/aggregation and dynamics. This feedback has been studied earlier in a real-case study with ICON by looking from the fixed-physics rather than the fixed-dynamics perspective. Our current results may indicate that sub-grid physics with stochastic physics perturbations emulate convective organisation effects.
\end{abstract}
\section{Introduction}



In the past three decades, skill scores of numerical models have greatly improved \citep[e.g.][]{bauer2015quiet}. A significant contribution to this quiet revolution was possible because of stochastic perturbation schemes \citep[e.g.][]{buizza1999stochastic,palmer2009stochastic,leutbecher2017stochastic}, especially to optimise ensemble spread. While this study focuses on deterministic physics in fixed dynamics experiments, our broader motivation also originates from the long-standing question about how deterministic and stochastic representation of sub-grid processes may differ. Therefore, we briefly review the significance of stochastic sub-grid physics first, framing the necessity to understand limits of deterministic suites. \\
In the integrated forecasting system (IFS) the Stochastically Perturbed Physics Tendency (SPPT) scheme leads to a good spread-skill relationship.  
It also enhances tropical variability in seasonal and decadal predictions. This includes improved representation of phenomena such as the Madden-Julian Oscillation (MJO) \citep{kessler2000rectification,vitart2010simulation,wcdstraus} and the El Niño-Southern Oscillation (ENSO) \citep{christensen2017stochastic,yang2019impact}. The SPPT methodology, originally developed for the IFS at ECMWF, has gained a lot of traction due to its simplicity and success. It is or has been part of many operational global and regional ensemble forecast systems across the world \citep{TowardRandomSamplingofModelErrorintheCanadianEnsemblePredictionSystem,wastl2019independent,lupo2020evaluation,frogner2022model,palmer2009stochastic,leutbecher2017stochastic}.
\\
In addition, more physically motivated stochastic schemes have been developed in recent years. Some examples describe stochastic representation of cumulus convection. In this representation multiple plumes and their interactions within a numerical weather prediction (NWP) model grid-box introduces sub-grid cloud population dynamics \citep{neggers2015exploring}. Several stochastic parameterisations of cloud production and dissipation have been explored within the NWP community \citep{hagos2018stochastic,neggers2015exploring,plant2012new}. 
\cite{bengtsson2011large,bengtsson2013,lisa2019,bengtsson2021} and \cite{bengtsson2024updates} have employed stochastic cellular automata to model convective organization and plume population dynamics. Notably, such schemes are integrated into the next-generation Global Forecast System (GFSv17) for operational global forecasting at NOAA. \cite{berner2017stochastic} and \cite{plantetal2015} offer comprehensive reviews of current operational stochastic parameterisation schemes.

Despite its large improvement of the spread-skill relationship in weather forecasts, little is known about the exact relation of deterministic versus stochastic physics with mesoscale atmospheric circulations. These circulations, at length scales below $\approx 1000$km, are often associated with convective cloud and precipitation systems, including mesoscale convective systems. 
They are simplified on a coarse grid, but become resolved at convection-permitting and finer grids \citep{groot2024,groot2023}. We analyse various deterministic physics suites under fixed dynamics. The first aim is to assess the general similarity of physics suites in the experiment. Another aim is, based on the deterministic suites, to investigate whether deterministic versus stochastic descriptions of \textit{model physics} can be connected with convective organisation and aggregation feedbacks of \cite{groot2024} and \cite{groot2023analysis}. 
We will assess whether these feedbacks may provide a physical explanations for the need of stochastic physics. \\

Here, we specifically investigate the response of precipitation and the physics tendencies of temperature and humidity to a common dynamical forcing (advection tendencies). Most dynamical meteorology students learn about the application of the omega equation and how it is derived \citep[e.g.][]{lackmann2011midlatitude,holton2013}. This equation tells us that geostrophic advection of temperature and vorticity are useful terms to identify where we expect upward motion, and hence, under sufficiently moist conditions, where precipitation is expected at mid-latitudes. Similarly, in the tropics, equatorial waves like Kelvin Waves are well-known as local propagator of the large-scale circulation, which co-determine the statistical distribution of convective systems near the inter-tropical convergence zone \citep[e.g.][]{knippertz2022intricacies}. However, the exact importance of dynamics for precipitation is not easy to quantify. We carry out a fixed-dynamics experiment with different single-column models (SCMs) as part of our Model Uncertainty Model Intercomparison Project (MUMIP; Christensen et al., in preparation; \cite{christensen2020}). In this project, we force the physics suites in SCMs with a DYAMOND benchmark simulation that partially resolves deep convection. Hence, while being exposed to a common dynamical evolution, our benchmark resolves more processes explicitly and parameterises fewer processes than our SCMs. We will be able to determine to which extent dynamical constraints dominate the uncertainty of precipitation predictions spanning the Indian Ocean for one month.
We compare precipitation predictions of our physics suites in deterministic mode with the deterministic 
benchmark.

\begin{figure}[htbp]
    \centering
    \includegraphics[width=100mm]{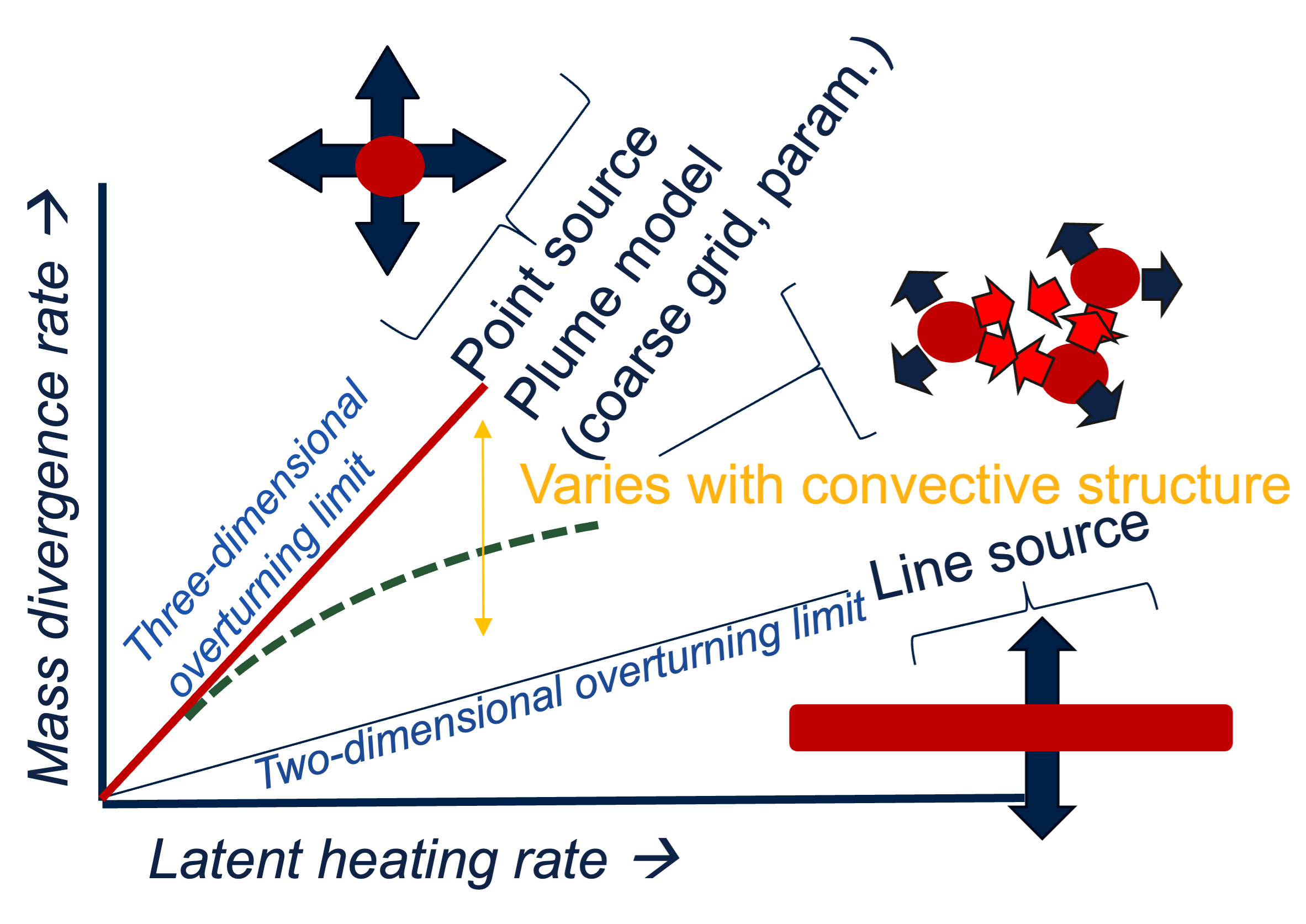}
    \caption{Under conditions of fixed column precipitation rate (as proxy for latent heating rate), the mass divergence rate depends on convective organisation and aggregation in large-eddy and convection-permitting ICON simulations, but not in coarser ICON simulations of the same case study \citep{groot2024,groot2023}; adapted from \cite{groot2024}, Figure 1. }
    \label{fig:conceptual}
\end{figure}

We hypothesise based on \cite{groot2024,groot2023} and \cite{groot2023analysis} (Chapters 6 and 7) that the coarse-grid physics suites 
are notably underdispersed in the (convective) precipitation when they inherit common dynamics from a benchmark (when compared with the variation between any SCM and the benchmark). 
%
This is because convective aggregation and geometry co-determine the outflow rate per unit precipitation rate 
in the benchmark, but that this quantity is relatively invariant at coarse-grids, i.e., for our physics suites (as for 13~km ICON, see \cite{groot2024}). In our experiment we have fixed the divergence-convergence pattern based on a DYAMOND simulation by definition, hence convectively initiated precipitation should be strongly constrained in the SCMs. We can quantify the strength of this constraint by correlating precipitation between physics suites. Under strongly constrained 
precipitation variability in the SCMs, substantial precipitation variability may still exist in convection-permitting simulations with equivalent large-scale dynamics
. This conditional variability may exist because fine convection-permitting models can resolve mesoscale circulations associated with convective organisation and aggregation, which are hypothesised to live by the virtue of gravity wave (bore) interactions \citep[see for instance][
]{groot2024,groot2023analysis,bretherton1989,npc1991,mapes1993}. The waves also result in mesoscale pressure fluctuations and they modulate the mesoscale vertical motion and convective initation \citep[e.g.][]{houze2004,adams2013,morrison2016,adams2020impact,kuo2025}. 
In addition to simply correlating precipitation between the physics suites, we can connect the structure of the joint precipitation PDFs with the physics tendencies and proxies of convective organisation and aggregation. As a result of the conditional variation of the dynamics feedback with convective organisation and aggregation, the best fit between predicted coarse-grid and predicted convection-permitting precipitation is expected to be slightly off-linear \citep[see ][ and Figure \ref{fig:conceptual}, green dashed line]{groot2024}. The difference between coarse-grid and convection-permitting preciptiation is also expected to correlate with convective organisation, and environmental proxies thereof. This is consistent with \cite{groot2024} and \cite{groot2023analysis}, but approaches the same problem in a reverse direction: rather than looking at ``fixed'' latent heating from convection, we here observe physics variation under fixed dynamical constraints.

\color{green}

\color{blue}
\color{black}
In Section \ref{methods} we describe our simulation data and diagnostics. In Section \ref{results} we assess the consistency of the data with our hypotheses. Finally, in Section \ref{summarydiscussion}, we summarise the findings and discuss them in the scope of stochastic perturbation schemes.
\section{Methods}
\label{methods}
\subsection{Model suites and single-column models}
We utilise the Model Uncertainty Model Intercomparison Project (MUMIP) datasets, which have been derived from the final 31 days of a 2.5 km DYAMOND ICON simulation of August-September 2016. We have coarse-grained the output of this simulation and constructed SCM forcing files for an Indian Ocean subdomain, following \cite{christensen2020}. 
Thus, in our single-column models (SCMs), we enforce the fixed dynamics to compute the corresponding physics (and total) responses of the following physics suites: \begin{itemize}
    \item IFS, cy48r1 \citep{81370}
    \item ARPEGE, version 6.4.2 \citep{roehrig2020cnrm}
    \item GFS, version 16 \citep{GFSv16,CommonCommunityPhysicsPackageFosteringCollaborativeDevelopmentinPhysicalParameterizationsandSuites,bengtsson2024updates}
    \item Rapid Refresh (RAP) \citep{RAP,CommonCommunityPhysicsPackageFosteringCollaborativeDevelopmentinPhysicalParameterizationsandSuites,RAPphysics}
\end{itemize}
We cover the entire month by integrating the SCMs out to six hours lead time over 44.000 grid boxes (22km spacing) 
for 31 days and 8 initialisations per day (except for the final day, with fewer initialisations). 
We extract forecasts of each SCM at lead times of 3hr to 6hr. Advection tendencies have been coarse-grained from the benchmark and we add up vertical and horizontal advection tendencies of ICON state variables to force SCM dynamics \citep[see][]{christensen2020,christensenetalinprep}. This implies that advection of hydrometeors like cloud ice and cloud liquid water is neglected.\\
The surface boundary is forced with either SSTs or prescribed surface fluxes, which depends on the choice of each modeling center. 
For more detailed information about the model suites we have utilised, we refer \cite{christensenetalinprep}.
\subsection{Diagnostics}
Our diagnostics for each column are averaged from 3 to 6hrs lead time (see Table 1). The mixed-layer (ML) tendencies represent a mass-weighted layer average over 0-500 m and the free-tropospheric (FT) tendencies represent a mass-weighted layer average between 500 and 16500 m. 
\begin{table}[htbp]
\caption{Definition and abbreviation of each of our diagnostics.}
\label{table-defs}
\begin{tabular}{|l|l|}
\hline
\textbf{Definition}                                                                                                         & \textbf{Abbreviation}    \\ \hline
Precipitation rate                                                                                                                       &- \\ \hline
\begin{tabular}[c]{@{}l@{}}Vertical mean tendency of specific humidity from model physics in the ML\end{tabular}                & MeteosHuML    \\ \hline
\begin{tabular}[c]{@{}l@{}}Vertical mean tendency of specific humidity from model physics in the FT\end{tabular}                & MeteosHuFT   \\ \hline
\begin{tabular}[c]{@{}l@{}}Vertical mean tendency of temperature from model physics in the ML\end{tabular}       & MeteoTeML      \\ \hline
\begin{tabular}[c]{@{}l@{}}Vertical mean tendency of temperature from  model physics in the FT\end{tabular}                       & MeteoTeFT   \\ \hline
\begin{tabular}[c]{@{}l@{}}U and V components of the wind speed (model levels)\end{tabular}                          & u, v                                \\ \hline
\end{tabular}
\end{table}\\
Since our benchmark does not have the same separation 
between physics and dynamics as the physics suites in SCMs (with 22~km grid spacing), we would expect physics and dynamics tendencies to differ substantially. This is mainly because the benchmark partially resolves deep convection, while this process is represented by a sub-grid parameterisation in SCMs. Unfortunately, the corresponding tendencies for the benchmark are also not directly available. Therefore, we reconstruct pseudo-ICON tendencies based on the dynamics we enforced in SCMs. In principle, each SCM could be used as reference dynamics. We reconstruct pseudo-ICON tendencies based on both IFS and ARPEGE and find virtually identical results; we select IFS-based ICON tendency reconstruction as default. Based on the total three-hourly change of ICON state variables we can thus reconstruct pseudo-physics tendencies by calculating the residual of total change and dynamics of an SCM. These pseudo-physics tendencies would have been valid if ICON had the same separation between explicitly resolved dynamics and physics as the coarse-grid models. 
That means, if the sub-grid parameterisations of our physics suites were a perfect fit to our benchmark, the SCM tendencies and their pseudo-tendency counterparts were equal. Hence, physics tendencies and pseudo-physics tendencies are comparable counterparts. The following pseudo-tendencies are extracted from our benchmark (with their SCM tendency counterpart abbreviated in brackets):
\begin{itemize}
    \item Vertical mean pseudo-tendency of specific humidity from model physics in the ML (MeteosHuML)
    \item Vertical mean pseud-tendency of specific humidity from model physics in the FT (MeteosHuFT)
    \item Vertical mean pseudo-tendency of temperature from model physics in the ML (MeteoTeML)
    \item Vertical mean pseudo-tendency of temperature from model physics in the FT (MeteoTeFT)
\end{itemize}

\subsection{Verification of MUMIP datasets}
It has been verified for the project that initial conditions and imposed dynamics are virtually identical for all physics suites. For a few samples this cannot be asserted, which is apparently a recurrent daily issue in the forcing files for one physics suite, and we remove these columns from the output. 
The output of the experiment is subset to the columns that have been completed correctly in a model pair or a single physics suite (when compared to the benchmark). The available number of columns per model varies from 87\% to 100\%, i.e., 
7 to 8 out of 8 initialisations per day and 9-11 million columns.

\section{Results}
\label{results}
\subsection{Joint precipitation PDFs}
\label{jointprecpdfs}
\begin{figure}[htbp]
    \centering
    \includegraphics[width=75mm]{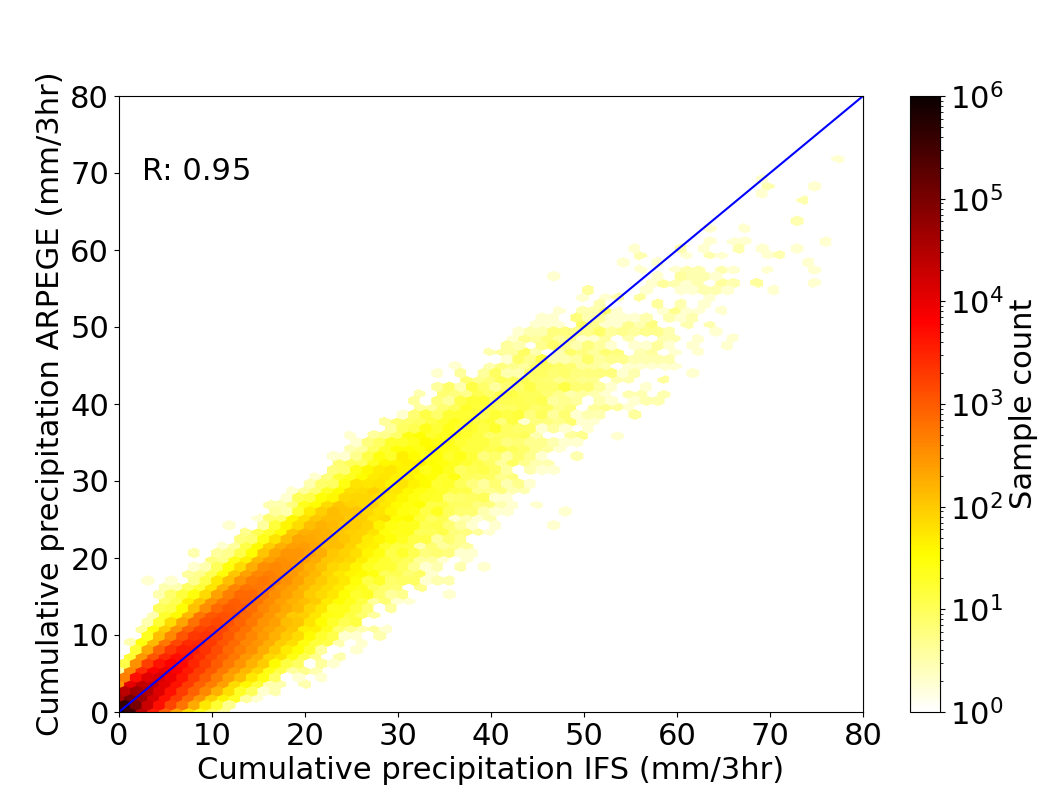}
    \includegraphics[width=75mm]{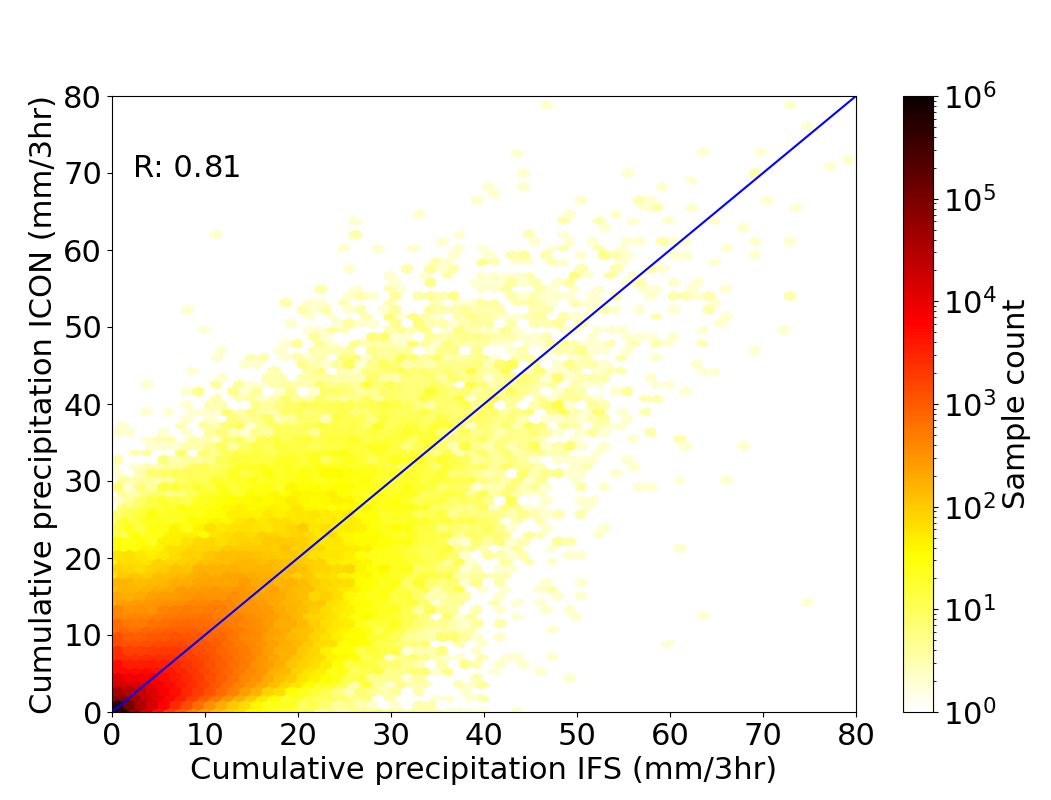}\\
    \includegraphics[width=75mm]{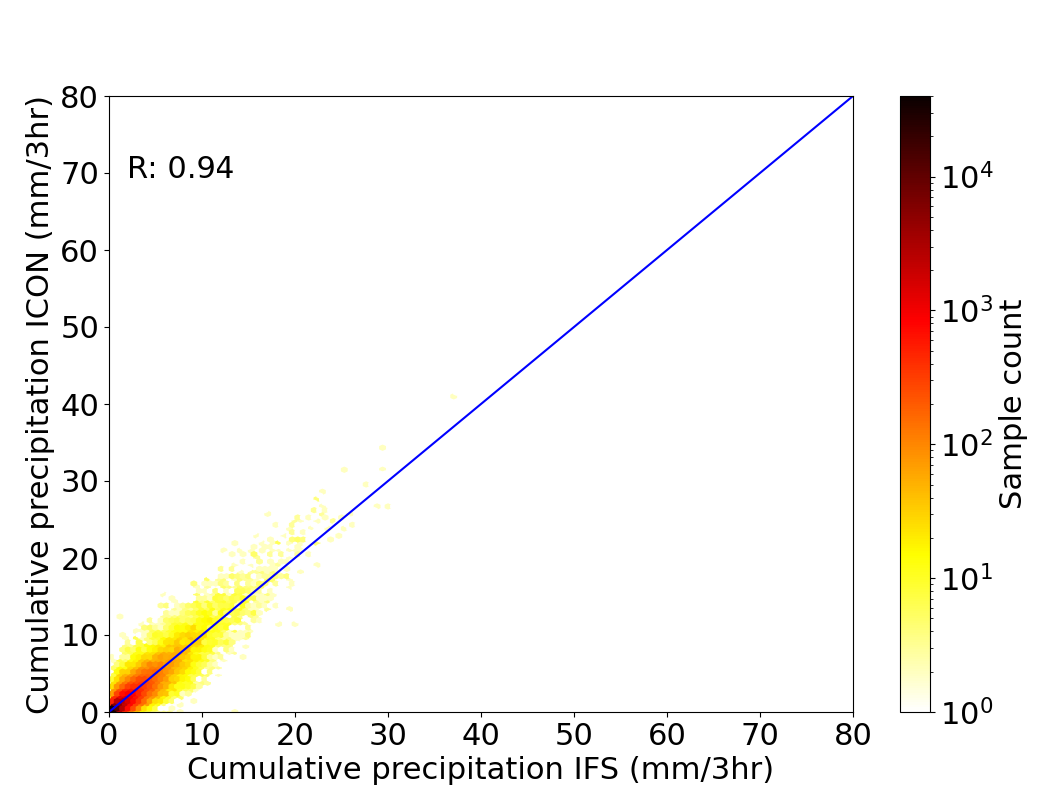}
    \caption{Joint PDFs of precipitation accumulation over all available columns, left top: for ARPEGE and IFS at lead times of 3hr to 6hr; right top: IFS (same lead time) and ICON over all available columns; bottom: same as right top, after regridding to 1.0 degrees. Blue lines represent the 1:1-relationship. 
    }
    \label{fig:JointPFDs}
\end{figure}

The joint PDF of the precipitation rate of IFS and ARPEGE is shown in Figure \ref{fig:JointPFDs} (top left panel). The plot shows that the vast majority of columns have precipitation accumulation very close to the 1:1-line. 
Typically, however, IFS slightly exceeds ARPEGE. The correlation between the models is 0.95 for column precipitation. 
Similar patterns are found by comparing any other pair of our physics suites, with correlation coefficients of 0.96-0.98. Interestingly, the correlation between IFS and ARPEGE is worse than between either of these and the RAP and GFS physics suites. Note that all models have a comparable mean precipitation rate of about 0.4-0.5 mm/3hr (after spin-up).\\
Figure \ref{fig:JointPFDs} shows the joint precipitation PDF of the high-resolution ICON and IFS at the native 0.2 degree grid (top right panel) and after coarse-graining to 1.0 degrees (bottom panel). The correlation between IFS columns and our benchmark drops to 0.81 for IFS (0.79 for ARPEGE). 
\\
The high precipitation spread of ICON for a given IFS precipitation rate, compared to spread between other pairs, could be explained by various factors. One of them is the active horizontal advection of hydrometeors like cloud ice and cloud liquid water in the benchmark, which, on the contrary, are not advected actively in the SCMs, following from the column assumption. The hydrometeor advection in the benchmark leads to net transport of cloud water away from the updraft cores, which tend to overlap with precipitation maxima. Therefore, hydrometeor advection is expected to smoothen out precipitation maxima when hydrometeors are advected. However, the resulting exchange of water will only occur between columns that are near neighbours. Furthermore, it is expected that the advection away from updraft cores is approximately proportional to the local precipitation rate and reduces the maximum. In-depth analysis of damping of precipitation maxima is, however, most relevant when comparing SCMs with multiple convection-permitting benchmarks, which is beyond the scope of this study. For a more in-depth discussion of hydrometeor advection and near-grid-scale diffusion, we refer to \cite{lisasanderfilipvana}.\\
Secondly, another source of increased precipitation uncertainty in the benchmark may stem from the resolved gravity-wave interactions in the convection-permitting benchmark. Fine-grid simulations are able to resolve the spectral interactions of gravity waves, whereas coarse-grid simulations may resolve the coarsest gravity wave mode only \citep[][and Figure \ref{fig:conceptual}]{groot2024,groot2023analysis}. On the other hand, at fine grids, spectral interactions then apparently reduce the conditional outflow rate under conditions when gravity waves can interact \citep[see][]{groot2023,groot2024}. Under fixed dynamical constraints, both the hydrometeor and gravity wave interactions may lead to enhanced precipitation spread in fine-grid simulations. \\
If we regrid our precipitation from $\approx 22$ km to $\approx$ 110 km (1.0 degree), hydrometeor advection becomes a negligible source of noise, because the average distance between columns increases and, consequently, local variability rapidly becomes under-resolved. Furthermore, purely as a result of horizontal hydrometeor advection, we only expect a linear proportionality of precipitation between the benchmark and our physics suites, which translates to a linear fit with negligible intercept as best fit. 
The expected best fit becomes slightly non-linear at these length scales if, however, the gravity wave interactions contribute as a substantial source of uncertainty to Figure \ref{fig:JointPFDs}, based on \cite{groot2024} and \cite{groot2023}. Then, a power relation would fit the joint precipitation PDFs between physics suites and the benchmark better (similar to green dashed line of Figure \ref{fig:conceptual}).  If $p > 1$, then gravity-wave interactions plausibly affect precipitation variability. If, however, $p=1$ then hydrometeor advection may be the only dominating difference between the benchmark and our SCMs for deep-convective precipitation. 
Finally, additional unidentified factors may also impact the joint precipitation PDFs.\\ 
When we coarse-grain the precipitation to a 1.0 degree grid representation, the pairwise correlation of the SCMs with our benchmark increases to 0.93-0.94 (Figure \ref{fig:JointPFDs}, bottom panel). However, pairwise correlations between the physics suites also increases to 0.98-0.99 after coarse-graining (not shown). Therefore, the joint PDFs of precipitation are so far consistent with our hypotheses: the effective relationship between precipitation in physics suites is purely linear and the relationship of each of them with the high-resolution, convection-permitting benchmark has considerable conditional spread, some of which may be explained by convective organisation and aggregation. 
Since substantial contributions to conditional spread of ICON remains at scales of about 100 km ( $\approx$ 1.0 degrees), this spread may indeed partially come from interactions of convective gravity wave perturbations. \\
We now test whether gravity wave interactions from well-organised convective systems \citep{groot2024,groot2023} are a plausible source of additional spread between coarse-grids and the benchmark. To test this, we fit the best models of predicted ICON precipitation based on accumulation in the SCMs. We thereby minimise mean-squared error of the precipitation prediction represented at the 1.0 degrees grid. We fit an exponential curve:
\begin{equation}
    y = a + b x^p
\end{equation}
Here, $x$ is the precipitation of an SCM, $y$ is ICON precipitation, $p$ is the exponent of interest, $a$ is an intercept with expectation very close to 0 and $b$ is a linear coefficient representing a multiplicative bias between different models when $a = 0$ and $p = 1$.\\
\begin{table}[]
\caption{Exponents $p$ of best fit between ICON precipitation and any SCM with their uncertainty based on 500 bootstrapped fits.}
\begin{tabular}{|l|l|l|l|l|}
\hline
SCM/Physics suite & Best estimate of $p$ & Uncertainty $p$ ($1{\sigma}$) & Best estimate of $a$ & Uncertainty $a$ ($1{\sigma}$)\\ \hline
IFS/IFS    & 1.06                               & 0.004 & -0.036 & 0.002                                                                                        \\ \hline
ARPEGE/ARPEGE & 1.05                               & 0.004   & 0.019 & 0.002                                                                                      \\ \hline
CCPP/GFS    & 1.02                               & 0.004  & -0.104 & 0.003                                                                                       \\ \hline
CCPP/RAP    & 1.05                               & 0.005   & -0.069 & 0.002                                                                                      \\ \hline
\end{tabular}
\label{exponentialfits}
\end{table}
As expected, $a$ is small. Values for $a$ and $p$ are tabulated in Table \ref{exponentialfits}. An uncertainty of $p$ is obtained from 500 bootstrap fits, estimating all parameters at once. Furthermore, the intercepts are very small: up to a 0.1 mm/3hrs absolute value, for GFS. As a result, tabulated exponents are insensitive to inclusion of an intercept value. However, GFS has a drizzle problem in our dataset, which means the model is rarely dry. The first percentile of precipitation is at 0.004 mm and fifth percentile 0.02 mm precipitation within our domain, which includes subtropics. We hypothesise that the poor fit of $a$ in GFSv16 is due to its strict convective initiation criteria which has been substantially revised in the upcoming coupled GFSv17 model \citep{bengtsson2024updates}. Therefore, we expect the GFS fit, which has the lowest exponent, to be less indicative of the targeted deep-convective organisation signals than that of the other physics suites. Each of the other three physics suites has an exponent of 1.05-1.06, which exceeds $p=1$ by about 0.05-0.06. Since the bootstrap uncertainty is at least an order of magnitude smaller (Table \ref{exponentialfits}) for these three suites (only about $p \approx 1+5\sigma_{p}$ instead for GFS), such robust fits are unlikely to arise randomly. It suggests statistical significance.\\
We can also remove columns with little to no precipitation from our dataset and repeat the fitting, which biases our fit towards deep convection. However, this would introduce two arbitrary decisions: 
\begin{itemize}
    \item Based on which model(s) do we reject samples?
    \item For which threshold do we subset our columns?
\end{itemize}
Nevertheless, it turns out that ARPEGE and IFS have a robust positive exponent, regardless of the arbitrary decisions above, whereby IFS has a robust value between 1.03 and 1.10 under reasonable constraints. The signal is less clear for RAP and, even more so, the GFS fit at high precipitation rates is highly dependent on arbitrary choices. Exponents are generally close to 1 for GFS (see also Table \ref{exponentialfits}).  \\
It is possible that GFS has a slightly different interaction between physics and dynamics in our experiment, but given the obtained correlations between all our physics suites, large deviations from other physics suites in precipitation-dynamics coupling is unlikely.\\
We do identify two relations in this subsection. Firstly,  we find  very high similarity in precipitation between physics suites. Secondly, we find our hypothesised signal of an off-linear feedback of precipitation to the dynamics in a convection-permitting model, but this signal is not so strong. This signal strength is expected to vary between convection-permitting models and regions, and, hence, should be made more robust with a future MUMIP experiment utilising additional convection-permitting models. A rather weak signal is expected for a model like ICON, which is known to have a lower degree of convective organisation compared to other models \citep[e.g.][]{christensen2021fractal,lilli,mooers2023,groot2024}. It has been challenging to assess the effect of geometry of convective cells even in a highly organised mid-latitude case \citep{groot2024} and the MUMIP dataset indicates even weaker signals. Nevertheless, in summary, a slightly non-linear feedback between physics and dynamics appears statistically robust in our benchmark and a purely linear feedback in coarse-grid models is very robust. \\
In the next subsections we move our main focus from precipitation to its relation with the model physics tendencies, and to the represented level of uncertainties between physics suites in these tendencies. Then, we move on to the relation of precipitation with the associated environmental wind shear in Section \ref{shearcomposits}. 
\subsection{Specific humidity physics tendencies in the free troposphere}
 We now consider the free-tropospheric specific humidity physics tendencies, MeteosHuFT (see Table \ref{table-defs}). These are strongly correlated with the precipitation rate (e.g. correlation of -0.97 in the IFS). This relationship 
indicates that precipitation corresponds to the vertically averaged water vapour sink. Small deviations from one to one correlations occur because of local vertical exchange with the mixed layer and, to a lesser extent, the stratosphere. \\
\begin{figure}[htb]
    \centering
    \includegraphics[width=90mm]{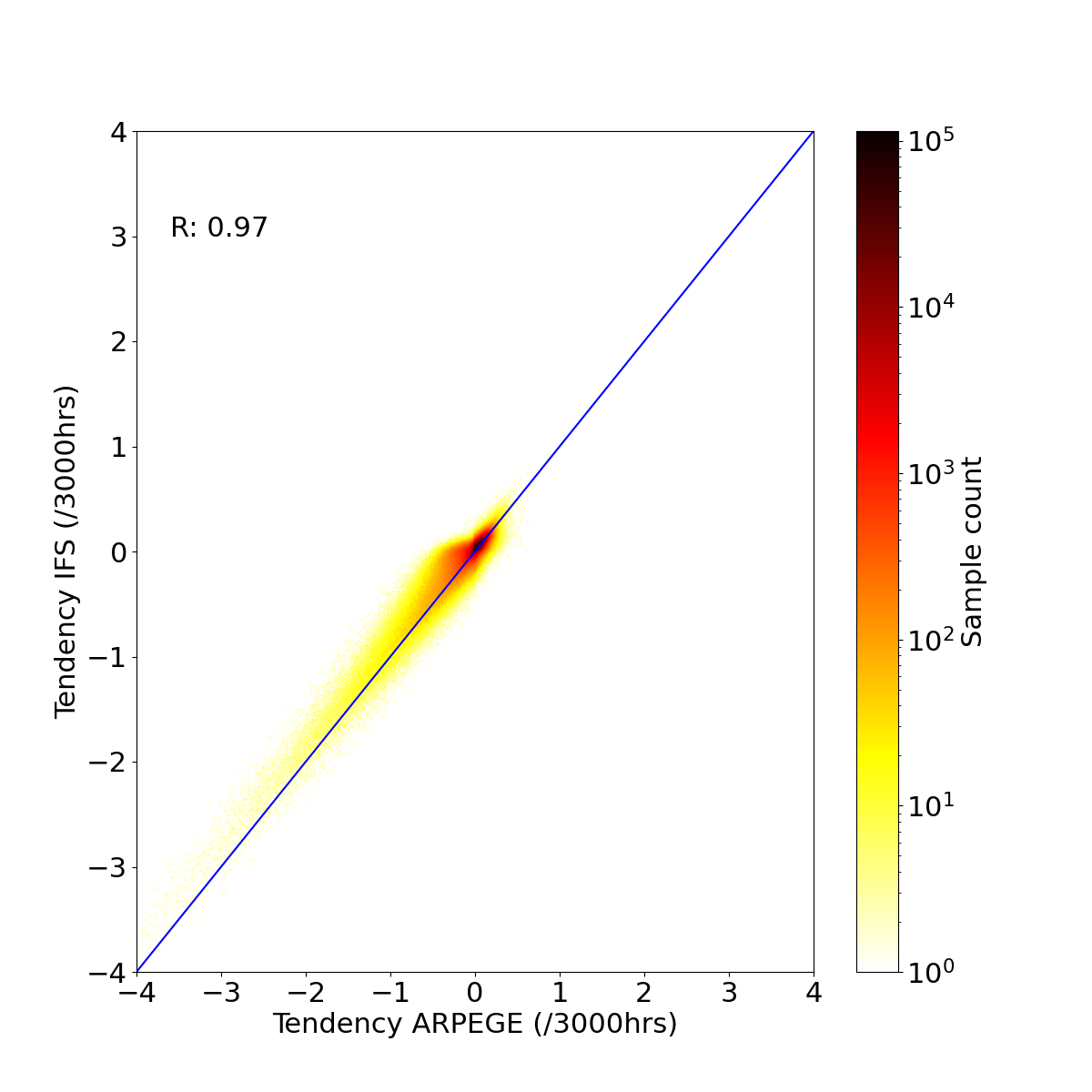}
    \caption{Left: Joint PDF of IFS and ARPEGE specific humidity tendency from the model physics between 3hr and 6hr lead time over all available columns, averaged over the free troposphere. The blue line represents the 1:1-relationship. 
    \color{black}}
    \label{fig:IFS-ARPEGE-precip-FT-hum}
\end{figure}
\color{purple}\color{black}
Similarly to precipitation, the water vapour sink correlates extremely strongly between corresponding columns for pairs of physics suites, that is, 0.97 between IFS and ARPEGE (Figure \ref{fig:IFS-ARPEGE-precip-FT-hum}).  \\
Now we move our focus to the correlation between MeteosHuFT of IFS and its pseudo-physics-tendencies counterpart of our benchmark, which is shown in Table \ref{FTMLcorrs}. 
We find a correlation of only 0.80 based on IFS dynamics output 
(with a pseudo-tendency obtained from ARPEGE dynamics: 0.81). Similarly, the correlation between MeteosHuFT of ARPEGE and the counterpart in our benchmark is 0.81 (0.82 with ARPEGE dynamics). Finally, similar to precipitation correlations (see Sect. \ref{jointprecpdfs}), the variability of each of the former correlations is small when replacing ARPEGE or IFS physics suites with GFS or RAP: within $\pm 0.03$ (not shown). 
\\
The large reduction of the correlation coefficient, when moving from SCM physics pairs to a comparison between any SCM physics and the benchmark, indicates a substantial source of missing sub-grid and sur-grid variablity in the deterministic physics suites. Here, sur-grid indicates variability at approximately the scale of the grid spacing, which is below the effective model resolution. In the remaining, we imply both sur-grid and sub-grid variability with the term sub-grid variability.
While we find a strong extra source of uncertainty in the benchmark when comparing their magnitude, the corresponding reduction in the correlation was also observed in the previous section with the joint precipitation PDFs (see also Figure \ref{fig:JointPFDs}). This extra source of uncertainty would substantially affect model dispersion in a full three-dimensional earth-system setting with the physics suites. \\
Furthermore, we have diagnosed that the physics tendencies of water vapour essentially integrate to precipitation fluxes. The considerable source of uncertainty remains unrepresented in the deterministic SCMs and, hence, physics suites, even at a 1.0 degree grid (see Section \ref{jointprecpdfs} and Figure \ref{fig:JointPFDs}, bottom panel). As a consequence, a plausible source of conditional spread in our benchmark with respect to SCMs is the non-linear dynamics feedback from sub-grid convective organisation and aggregation, as suggested in the previous section. 


In summary, we observe considerable under-dispersiveness in the column precipitation and the associated vertically averaged physics tendency of specific humidity in SCMs compared our benchmark. We suggest that unresolved variability, which is constrained by the grid spacing, explains the underdispersion, following a pattern that was expected from findings of \citep{groot2024}. The under-dispersion is present in the free-tropospheric mean specific humidity tendencies of physics suites, as these are tightly correlated with precipitation accumulation. Moreover, multi-model ensembles of deterministic coarse-grid physics do not appear to represent the full range of diversity of mesoscale variability of a convection-permitting simulation, such as our benchmark, and are extremely unlikely to span all realistic uncertainty sources from sub-grid processes, including convective variability. \\
We will further assess the relation between the joint precipitation PDFs and the convective environment in Section \ref{shearcomposits}. 
In the next subsection, however, we address the correlations of mixed-layer-physics tendencies and of the free-troposphere (temperature only) between our physics suites, respectively, between physics suites and our benchmark. 
\subsection{Variability of other vertically averaged physics tendencies
}
We next consider the remaining vertically averaged physics tendencies, starting with temperature in the free troposphere. Table \ref{FTMLcorrs} shows that, in the free troposphere, the correlation between IFS and ARPEGE SCMs is extremely high at 0.97. The benchmark and SCM temperature tendencies are also much more tightly associated than the humidity tendencies, with correlations of 0.93 and 0.94. As for specific humidity physics tendencies, all physics suites correlate better with each other than each does with the benchmark. The variability among the four suites for this diagnostic remains within $\pm 0.02$ (not shown). Since the reduction in correlation is less pronounced for temperature tendencies than humidity, this suggests less under-dispersiveness for temperature. 
This result does not mean that stochastic physics would necessarily be unimportant for physics tendencies of temperature in the free troposphere, but it suggests humidity may be the dominant unrepresented sub-grid uncertainty among SCM physics tendencies.   \\
Moving on to mixed-layer tendencies, physics suites of IFS and ICON-DYAMOND are much less well correlated: on the order of 0.5 (Table \ref{FTMLcorrs}). If we compare this correlation to correlations between pairs of physics suites, it is notable that the physics humidity tendencies between ARPEGE and IFS in the mixed layer are also weakly correlated. Furthermore, Table \ref{FTMLcorrs} suggests that SCMs are only slightly overconfident; the difference in correlation between pairs of physics suites and any physics suite with ICON may be considered limited. \\
While being negligible for free-tropospheric uncertainty, surface coupling becomes important for uncertainty in the boundary layer. We use fixed prescribed SSTs in our ARPEGE and IFS simulations, whereas the CCPP simulations with the GFS and RAP physics use prescribe surface fluxes \citep[see][]{christensenetalinprep}. These uncertainties will affect the mixed-layer correlation patterns. Accordingly, for the mixed layer the four physics suites cannot be regarded as fully fixed to prescribed dynamics; the dynamics could be perturbed rapidly by surface processes. 
The exact interpretation of correlation patterns in the mixed layer fall beyond the scope of this work, and moreover, they are strongly affected by adjustments of the humidity profiles during spin-up of the SCMs \citep{grootetalinprep}. Strong indications of under-dispersion of SCMs is not found, unlike in the free troposphere. The large uncertainties and low correlations in the mixed layer may suggest that the error range in the boundary layer has to be sampled decently with multiple physics suites and two different surface couplings options. Finally, one could question whether our benchmark and SCMs represent boundary-layer processes well enough for firm conclusions. 
\\

\begin{table}[]\caption{Correlation table of column-to-column layer-averaged tendencies from physics suites in the MUMIP dataset. ML: mixed layer; FT: free troposphere. 
}\label{FTMLcorrs}
\begin{tabular}{|l|l|l|l|}
\hline
\textbf{Quantity, layer}   & \textbf{IFS-ARPEGE}         & \textbf{IFS-ICON}                  & \textbf{ARPEGE-ICON}               \\ \hline
Precipitation accumulation & {\color[HTML]{036400} 0.95} & {\color[HTML]{34FF34} 0.81}        & {\color[HTML]{34FF34} 0.79}        \\ \hline
Specific humidity, FT      & {\color[HTML]{036400} 0.97} & {\color[HTML]{34FF34} 0.80 (0.81)} & {\color[HTML]{34FF34} 0.81 (0.82)} \\ \hline
Temperature, FT            & {\color[HTML]{036400} 0.97} & {\color[HTML]{036400} 0.94 (0.94)} & {\color[HTML]{036400} 0.93 (0.93)} \\ \hline
Specific humidity, ML      & {\color[HTML]{F56B00} 0.45} & {\color[HTML]{F56B00} 0.42 (0.41)} & {\color[HTML]{CB0000} 0.22 (0.24)} \\ \hline
Temperature, ML            & {\color[HTML]{FFCC67} 0.58} & {\color[HTML]{FFCC67} 0.50 (0.51)} & {\color[HTML]{F56B00} 0.42 (0.47)} \\ \hline
\end{tabular}
\end{table}

\subsection{Conditional environments of strong precipitation biases}
\label{shearcomposits}


In the following, we address convective wind shear environments under varying ICON-IFS precipitation biases. Highly linear squall lines features are thought to co-vary with more precipitation in our benchmark than in the SCMs \citep{groot2024,groot2023analysis} under the fixed dynamical constraints. Strong low-level shear fosters linear squall line organisation \citep[e.g.][]{Lemone} and the direction of wind shear may be important for linear segments, their geometry, overturning and their orientation. For instance, quasi-two-dimensional overturning may be more favourable under directional shear between low-levels (below 800 or 900 hPa) and mid-levels (400-850 hPa) \citep{Lemone,Trieretal1997}. This alternative overturning is exactly what we try to separate from isolated convection, following \cite{groot2024}. Nevertheless, the exact relation between wind shear and long-lived, well-organised deep convection is still actively investigated, especially in the tropics \citep[e.g.][]{mark}.\\
As a result of the difference in overturning between isolated convection and highly-organised squall lines, their dynamics at a given precipitation rate differs \citep[][and the green dashed line of Figure \ref{fig:conceptual}]{groot2023,groot2024,groot2023analysis}. How much they differ is hypothesised to depend on the exact characteristics of overturning. Consequently, under fixed dynamics, linear, and to a lesser extent highly aggregated, convective features are more likely when coarse-grid models like SCMs have a negative bias; isolated convection is more likely when the SCMs have a positive precipitation bias compared to the benchmark. \\
To investigate the bias in our data, we investigate bins within the joint PDF of Figure \ref{fig:JointPFDs} (right top). Considerable precipitation is required in both the benchmark and a physics suite (here: IFS) to identify convective systems. Furthermore, and a precipitation ratio between models that below 0.5 or exceeding 2 is required for strong bias, because overturning characteristics alone may cause differences of ratios up to 2-3 \citep{groot2024,groot2023,npc1991}. 
As a control sample, we also take samples with nearly equal precipitation in IFS and the benchmark. We first select columns in a narrow bin that experience between 10 and 10.5 mm for IFS. From these, we subsequently isolate the columns with 2.85 to 3.00, 10.0 to 10.5 and 22.5 to 25.0 mm in the benchmark. 
For each of these three cases, we separately extract the wind profiles from the benchmark. We repeat this procedure for a bin between 7.0 and 7.5 mm in IFS, retaining the relative precipitation ratios between IFS and the benchmark. In the following, we will separately describe the obtained mean wind profiles of the subset with a low, neutral and high precipitation bias between the benchmark and IFS for each set. 
\\
We first compute a vertically averaged wind in the benchmark. It is obtained by weighing all model levels below 150 hPa equally. Model levels are denser at low levels and this meets our objective, because low levels are considered more important for convective inflow and organisation \citep[e.g.][]{Lemone,Trieretal1997,mark}. We define this vertically averaged wind to be the estimate of storm motion. We then plot the mean deviations of the wind at any height from this storm motion estimate. Across our six bins, with approximately 100-250 samples each, we compare this inflow estimate as a proxy for the overturning.\\ 
\begin{figure}
    \centering
    \includegraphics[width=75mm]{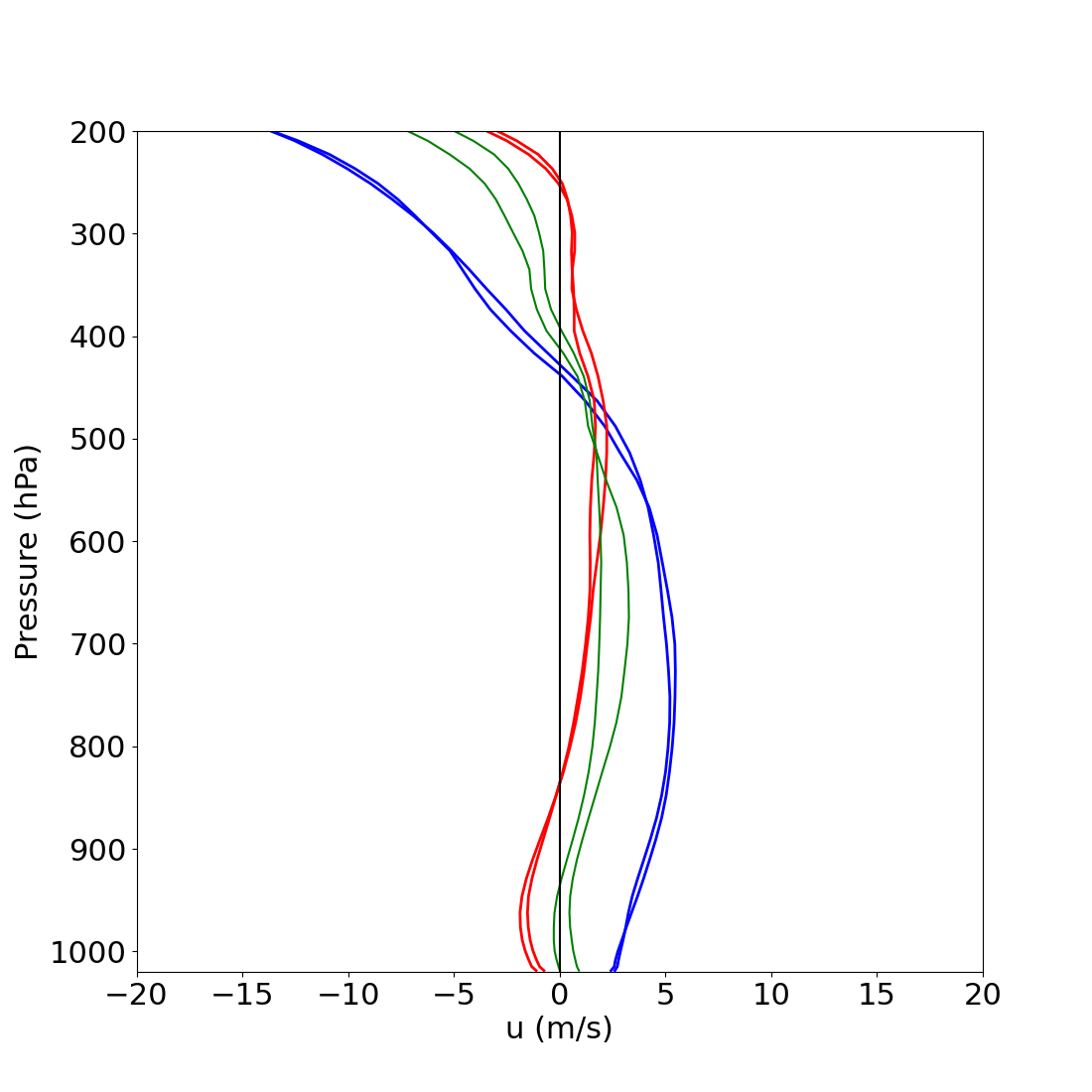}
    \includegraphics[width=75mm]{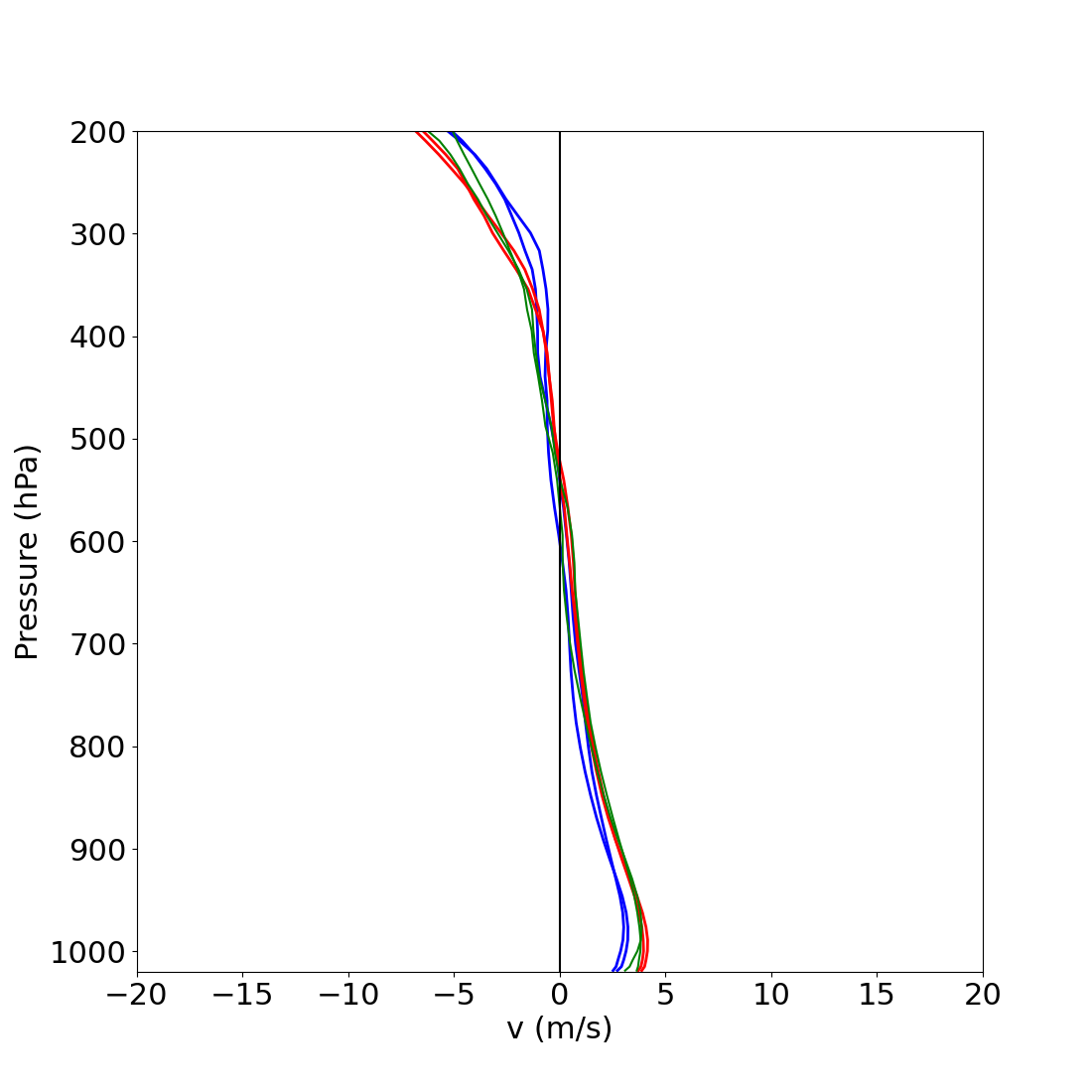}\\
    \includegraphics[width=75mm]{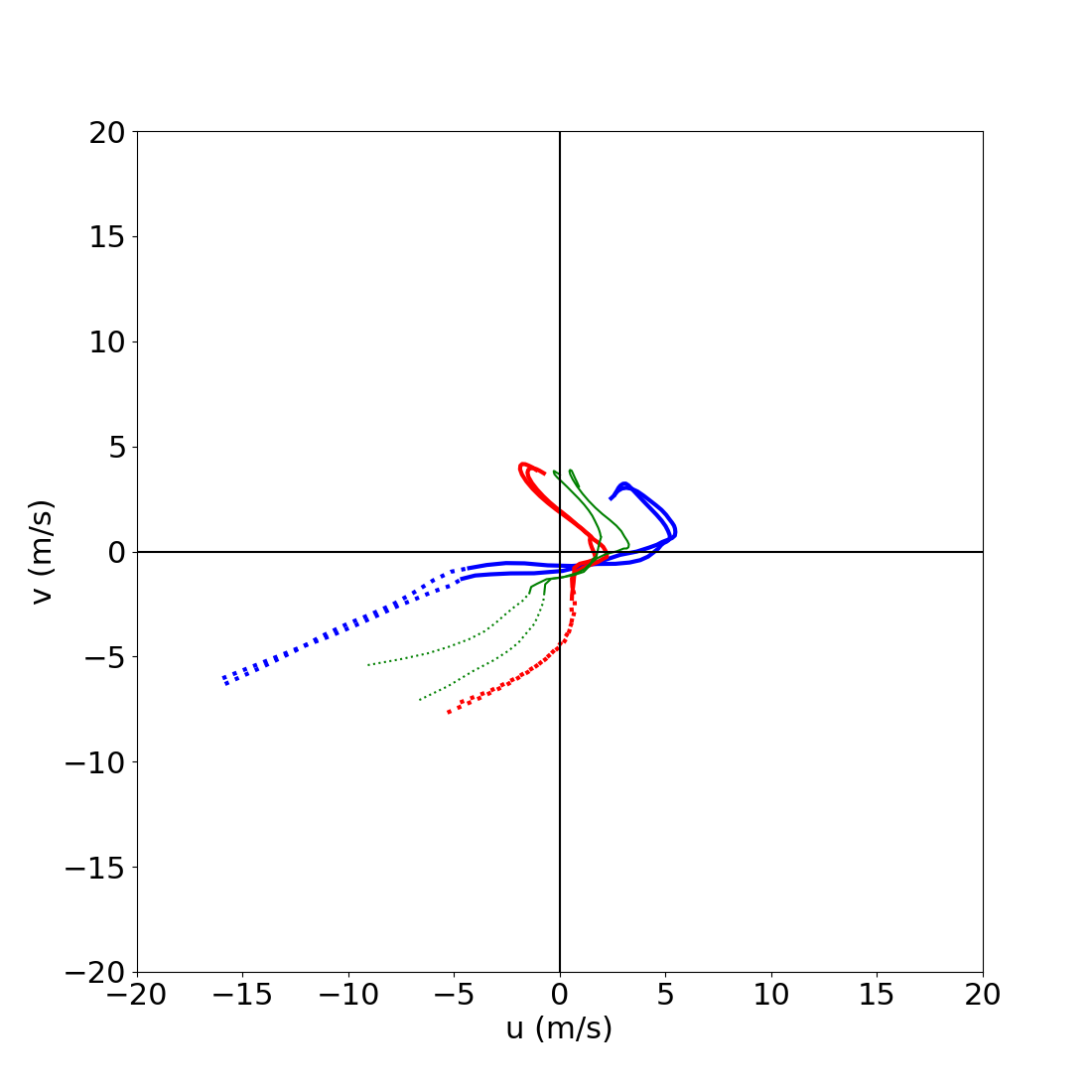}
    \caption{Top left: mean profile of u-component of the storm-relative wind. Top right: mean profile of the v-component of the storm-relative wind. Bottom: Composite mean hodograph in six bins of the joint precipitation PDF of Figure \ref{fig:JointPFDs} (top right) from surface to 350 hPa (solid) and up to 180 hPa (dotted). The thick blue lines indicate IFS has a positive bias with respect to ICON, the thin green lines indicate no bias and the thick red lines indicate a negative bias. 
    }
    \label{fig:hogograph}
\end{figure}
Figure \ref{fig:hogograph} shows the compsite u- and v-winds (top panels) and the hodographs of storm-relative winds (bottom) of our bins.
The thick blue hodograph, with negative precipitation bias in IFS, has the strongest deep shear, but it is mostly at uppermost levels. The low-level shear is comparable, if not weaker, than in the thin green and, especially, the thick red hodograph. The meridional shear looks similar in each subset, with most shear below 850 hPa. It is slightly weaker in blue profile (Figure \ref{fig:hogograph}). Most zonal shear of the green, and even more so, red profile occurs at 950 to 700 hPa. However, the blue profile has its zonal shear mostly above 550 hPa and below 900 hPa, the latter of which represents boundary layer friction. Furthermore, the curvature at low levels differs substantially from blue via green to red, favouring northerly inflow (red and green) as opposed easterly deep-shear-parallel inflow in (blue). In general, the blue sample has the weakest lateral deviations from the mean shear vector, which is almost zonally oriented. Finally, if we inspect the spread within each sample (not shown), the blue profiles also indicate generally less shear perpendicular to the shear vector. 
On the other hand, green and red hodographs indicate considerable lower-to-mid-level directional shear, comparable to the curvature of \cite{Trieretal1997,Lemone}. We would imagine inflow pattern differs between these  composite hodographs and they probably are associated with some variation in convective overturning. Curved inflow with a strong negative IFS bias fosters partial quasi-two-dimensional overturning, consistently with our hypothesis, and broadly consistent with analyses by \cite{Lemone,Trieretal1997,groot2024}. \\
In summary, we do find a relation between the wind shear at mid-levels, the estimated inflow orientation 
and the sign of the precipitation bias of IFS compared to ICON, supporting our hypothesis. 


\section{Synthesis}
\label{summarydiscussion}


\subsection{Summary}
Under fixed dynamical constraint of a 31 day Indian Ocean simulation, we have compared precipitation, physics tendencies and wind shear between convection-parametrised SCMs and a high-resolution convection-permitting benchmark. The dynamical constraints essentially fix SCM precipitation. Precipitation is accordingly highly correlated between SCMs ($>0.95$), which capture coarse-grid physics and dynamics. SCMs correlate much more weakly with our convection-permitting benchmark ($r \approx 0.8$), i.e., the convection permitting model has larger conditional uncertainty for fixed large-scale dynamics. 

Based on our analysis and \cite{groot2024,groot2023analysis}, two dominant factors may be responsible for this in our dataset: firstly, horizontal advection of cloud liquid and ice water and, secondly, gravity wave interactions which reduce the net outflow rate under improved convective organisation such as two-dimensional squall line overturning. While the first factor likely smoothens precipitation maxima, the second is likely to affect the mean precipitation bias between ICON and the MUMIP datasets over entire convective systems, when there is sufficient variation in convective organisation and aggregation within the benchmark. \\
Effects of the horizontal advection of hydrometeors may be effectively eliminated by analysing precipitation at a coarser 1.0 degree grid. On the resulting dataset, we test hypotheses to infer whether gravity wave interactions may play a critical role. It is expected that the optimal fit between precipitation in the benchmark and the SCMs increases more than linearly with precipitation rate. Indeed, we find robust deviations, with exponential relations minimising errors. However, the exponents is lower than between mass divergence rate and precipitation rate in \cite{groot2024}. Furthermore, while values are 1.05-1.06 for most physics suites, the exponent of GFS physics is only 1.02. 
A drizzle problem in the GFS physics simulations may significantly affect optimal fits. 

In deterministic SCMs, precipitation may effectively be seen as the exact water vapour sink from the free-troposphere. From our layer-averaged tendency analysis, it is apparent that SCMs are most overconfident about the averaged physics tendency of humidity, and overconfidence in temperature is also apparent. Future MUMIP experiments with other benchmarks could confirm overconfidence.\\ 
Given the relationship between convective organisation and the large-scale wind shear, we should expect different SCM precipitation biases to correspond to different convective wind shear environments. By compositing conditional shear profiles for two samples of neutral, positive and negative biases, we test this hypothesis and confirm that negative SCM biases are associated with shear profiles tending towards linear quasi-two-dimensional overturning. 

\subsection{Discussion}
\subsubsection{Complementary experimental design}
The SCMs are each hydrostatic models. However, they use prescribed dynamics from a non-hydrostatic model, in which vertical velocity is an explicit prognostic. Together with grid spacing, this is a key difference between convection-permitting and parameterised models. Both key differences also affect which horizontal dynamics is actively resolved and the tuning of parameterisations. However, an earlier study \citep{groot2024} has strongly suggested that the key differences for the mass divergence-precipitation relationship between convection-permitting and deep-convection parameterised simulations in ICON are \underline{\textit{not}} caused by switching off the parameterisation 
at a 13 km grid spacing 
\citep{groot2024}. 
Arguably, we can safely assume the horizontal grid spacing is critical in our evaluation. The former study also hypothesised that gravity wave interactions are critical. In our SCM experiments dynamical evolution is solely prescribed using boundary conditions derived from high-resolution ICON, which fixes the coarsest-mode gravity waves. On the contrary, it removes their critical variability at shorter wavelengths.\\
Statistically, we can say that our hypotheses based on \cite{groot2024} are supported by MUMIP data. Our analysis complements the real-case comparison between two versions of ICON in \cite{groot2024} and thus suggests generalisability of the differences between high-resolution ICON and coarse-grid models. Finally, we argue, it provides further support for their hypothesis that gravity wave interactions are likely a critical process altering convective feedbacks when increasing a model resolution from about 20 km to km-scale \citep{groot2024,groot2023,groot2023analysis}.
\subsubsection{Implications for representation of sub-grid variability}
As we move into the era of ERA5, with trained AI forecasting models such as AIFS and many others by commercial parties, we hugely depend on the best models, best datasets and their underlying assumptions \citep[e.g.][]{gencast,lang2024aifs,lang2024aifsens,lang2025multi}. This includes explicit and implicit assumptions made in our physics suites. For machine-learning-based weather and climate modelling based on ERA5 \citep{hersbach2020era5,soci2024era5}, we foremost depend on the characteristics of the Integrated Forecasting System (IFS) \citep{81370,81371,81368,81369}. Our results appear to provide a key component in a chain of evidence as to why deterministic physics may be overconfident. \\
The consequence of overconfidence for derived datasets that depend on a deterministic forecasting model with sub-grid parameterisations, such as a determistic ERA5, may be the following: 
\begin{itemize}
    \item dependence all the way down a compounded chain of data-driven models, ultimately issuing automatic products for warnings and wider applications \citep{sessionEMS}
    \item risk of overfitting, leading to sub-optimal minimisation
\end{itemize}
Compound training processes in a chain are typically not minimised in unison, which must imply \textit{training} minima will occassionally deviate from true minima, due to connection with some parent dataset(s). \\
Stochastic physics provides a solution to overcome the overconfidence in certain applications.

\color{black}
\subsubsection{Significance for stochastic perturbation schemes}
Essentially, we could consider Table \ref{FTMLcorrs} as an indicator of how similar model physics suites are under fixed dynamics. This is because precipitation is the main model physics process, which redistributes vast amounts of heat.
\\
We could invent an \textit{approximate} diagnostic for overconfidence of individual physics tendencies in the experiment. We would first define the unexplained variance of these tendencies as $u = 1-R^{2}$ of Table \ref{FTMLcorrs}. Then, we may estimate the ratio of this unexplained variance between pairs of physics suites over the unexplained variance between a physics suite and our benchmark as an overconfidence ratio: $O = \frac{u_{pair}}{u_{bg}}$. If our sample includes random physics suites with little correlation between them, this ratio will approach 1 for well-calibrated physics suites and optimal transferability. From the data in Table \ref{FTMLcorrs}, $O$ appears low in mixed-layer physics tendencies, which suggested little to no overconfidence. 
For humidity in the free-troposphere, $O\approx 6$. However, for temperature in the same layer, $O \approx 2$ or $3$.  
Because these tendencies drive significant precipitation, precipitation is also under-dispersed when compared between coarse-grid models. 
\\
Tailored stochastic perturbation schemes could overcome the large overconfidence. Our statistical links indicate why a scheme with properties of SPPT has great positive impact, if we slightly perturb the relation between physics humidity tendencies and precipitation. Stochastically perturbed parameterised tendencies (SPPT) in IFS was an operational scheme that could emulate unresolved convective organisation at mesoscales, because perturbations have the following properties:
\begin{itemize}
    \item \textbf{They perturb exact local water conservation and can emulate convective organisation}\\
    The specific humidity tendencies from the physics are adjusted stochastically in complete isolation from the fully deterministic water species and precipitation \citep{sarahjanelocketal2024}. 
    Hence, the conditional spread humidity tendencies from physics (Figure \ref{fig:IFS-ARPEGE-precip-FT-hum}) is increased. This mimics a perturbed humidity sink under fixed precipitation rate, which resembles modulating effects of the convective environment and structures, such as squall-line organisation. The latent heat release that the model dynamics experiences differs from the latent heating actually removed by the physics. 
    The devation of the adjustment would lead to a different amount of divergent outflow \citep[see also][Chapters 6 and 7]{groot2024,groot2023analysis}. 
    \item \textbf{They are of multiplicative nature}\\
    The multiplicative character of the stochastic perturbations leads to an increase of the mismatch between physics tendencies and precipitation rate with the increase of the expected precipitation rate of deterministic physics schemes. Thereby, it ensures a successful emulation of near-grid variability of convective structure. The noise factor can represents a pseudo-effect of convective organisation and aggregation. 
    Such effects evolve naturally at fine grids through gravity wave interactions \citep[see also][]{groot2023,groot2024,groot2023analysis} and goes at the cost of exact local water vapour conservation. When instead water conservation is imposed, we directly fix the precipitation (see Figure \ref{fig:IFS-ARPEGE-precip-FT-hum}) with strong overconfidence. 
    \item \textbf{They perturb \textit{at least} specific humidity}\\ 
    Because its overconfidence is suggested to be the largest, according to $O$ (using Table \ref{FTMLcorrs}).
    \item \textbf{They are tapered in both boundary layer and the stratosphere} \citep[e.g.][]{leutbecher2017stochastic}\\
    Because the differential uncertainty is very large in the free troposphere (Table \ref{FTMLcorrs}) and the layers to which SPPT is applied match perfectly with this strong under-dispersiveness.
\end{itemize}

It is plausible to assume that our physics humidity tendency signal is dominated by a lower free tropospheric source of moisture, for instance below 700 or 800 hPa. If we assume so, these results are neither necessarily consistent nor necessarily inconsistent with \cite{christensen2020}. This is because the study has shown that multiplicative nature of stochastic perturbations in SPPT holds better at lower levels than further up. Nevertheless, low-level sources of humidity 
represent only a dominant contribution to the deterministic humidity tendencies, which are perturbed by stochastic schemes.  \newline
Furthermore, if we think of the broader modelling world, we could hypothesise that the better a model reproduces highly organised squall lines, the more significant the off-linearity of joint precipitation PDFs of Figure \ref{fig:JointPFDs} is. Important examples of off-linearity likely include mid-latitude summer squall lines over continents \citep{groot2024}. Under conditions of highly organised convection with strong shear and squall line potential, reliability of stochastic physics could \textit{sometimes} be particularly important \citep[e.g.][]{rodwell2013characteristics,rodwell2018flow,rodwell2023uncertainty,groot2025}. Further investigation with a variety of benchmark simulations and/or regions is needed to give insight in the non-linear perturbation effect of quasi-two-dimensional squall-lines. \\
Lastly, explanatory power of stochastic physics should be sought in all sub-grid processes, without restrictions to deep convection and precipitation, although they may currently dominate under-represented sources that urge for stochastic perturbation. 
\section*{Acknowledgements}
EG and HC acknowledge funding from the Leverhulme Trust, Grant Number: RPG-2022-192. \\
XS and JS were supported in part by NOAA cooperative agreement NA22OAR4320151, for the Cooperative Institute for Earth System Research and Data Science (CIESRDS). The statements, findings, conclusions, and recommendations are those of the author(s) and do not necessarily reflect the views of NOAA or the U.S. Department of Commerce. \\
The research of LB is supported by US federally appropriated funds.\\ 
The authors would like to thank Ligia Bernardet for checking the original draft and her role in the initial development of the Model Uncertainty Model Intercomparison Project. 
\section*{Author contributions}
Conceptualisation, formal analysis and writing of original draft: EG; Project data curation: EG, HC, XS, KN, WL, RR; Reviewing: EG, HC, XS, WL, LB, JS; Funding acquisition: HC.
\\



\newpage
\color{black}
 \bibliographystyle{plainnat}
\bibliography{refs.bib}
\end{document}